\documentclass[twocolumn,showpacs,preprintnumbers,amsmath,amssymb]{revtex4}

\usepackage{graphicx}
\usepackage{dcolumn}

\usepackage{bm}
\def\Arcosh{\mathop{\rm Arcosh}\nolimits}

\begin{document}
\title{Critical adsorption on non-spherical colloidal particles}
\author{S. Kondrat, L. Harnau, and S. Dietrich}
\affiliation{
         Max-Planck-Institut f\"ur Metallforschung,  
         Heisenbergstr.\ 3, D-70569 Stuttgart, Germany, 
         \\
         and Institut f\"ur Theoretische und Angewandte Physik, 
         Universit\"at Stuttgart, 
         Pfaffenwaldring 57, 
         D-70569 Stuttgart, Germany
	 }
\date{\today}
\begin{abstract}
We consider a non-spherical colloidal particle immersed in a fluid close to 
its critical point. The temperature dependence of the corresponding order 
parameter profile is calculated explicitly. We perform a systematic expansion 
of the order parameter profile in powers of the local curvatures of the surface 
of the colloidal particle. This curvature expansion reduces to the short 
distance expansion of the order parameter profile in the case that the solvent 
is at the critical composition.
\end{abstract}
\pacs{64.60.Fr, 68.35.Rh, 82.70.Dd, 64.75.+g}
\maketitle

\section{Introduction}
Various interactions between colloidal particles immersed in a solvent 
can influence the properties of the system. In a colloidal suspension
containing particles of different size, there is an effective attractive interaction
between larger particles. This attraction is due to the extra volume that 
becomes available to smaller particles when larger particles approach each 
other, leading to an overlap of excluded volumes which increases the entropy 
of the system (see, e.g., Ref.~\cite{harn:07} and references therein). Colloidal 
particles, surfaces of which dissociate in solution, exhibit a screened Coulomb 
interaction due to charged surfaces of the particles and the surrounding counterions 
(see, e.g., Ref.~\cite{hans:02}). Van der Waals dispersion forces arise from induced 
dipole-dipole interactions due to quantum mechanical fluctuations of the charge 
density (see, e.g., Ref.~\cite{pars:06}). The confinement of critical order parameter 
fluctuations in a binary liquid mixture near its critical demixing point 
give rise to long-ranged critical Casimir forces between immersed colloidal 
particles (see, e.g., Ref.~\cite{schl:03}). The richness of the physical properties 
of colloidal suspensions is mainly based on the possibility to tune these interactions 
which differ significantly in strength and range. In the case of the effective entropic
interaction, the Coulomb interaction, and the van der Waals interaction this 
tuning is accomplished by changing the composition of the 
solvent by adding depletion agents, salt, or other components. For example,
by matching the indices of refraction of the colloidal particles and the 
solvent it is  possible to effectively switch off the dispersion forces and  
to create colloidal suspensions in which the actual effective interaction  
between uncharged colloidal particles very closely resembles a hard core 
potential. In the case of charged colloidal particles, even small polyelectrolyte
additives can have substantial impact on the aggregation and kinetic stability
of the charged particles \cite{harn:02}.
Compared with such modifications, changes of the temperature or 
pressure typically result only in minor changes of the effective interaction 
between colloidal particles. However, effective interactions generated by 
bringing solvents close to a phase transition of their own are extremely 
sensitive to such changes as observed experimentally
\cite{beys:85,gurf:89,gall:92a,gall:92b,nara:93,kurn:95,kurn:97,jaya:97,peti:98,law:98,beys:99}.
In these experiments, a dilute suspension of spherical colloidal particles 
has been formed within the one-phase region of a binary liquid mixture
acting as the solvent. As the temperature of the system approaches the 
phase separation temperature of the binary liquid mixture, 
the colloidal particles aggregate and flocculate out of 
solution. Light scattering measurements indicate that an adsorption layer 
rich in one of the two solvent species forms around each colloidal particle.
The experimental results strongly suggest that the colloidal aggregation
behavior is induced by the presence of this adsorption layer. The thickness
of the layer increases upon approaching the aggregation line in the phase diagram.

\begin{figure}
\begin{center}
\includegraphics[ width=6cm,clip]{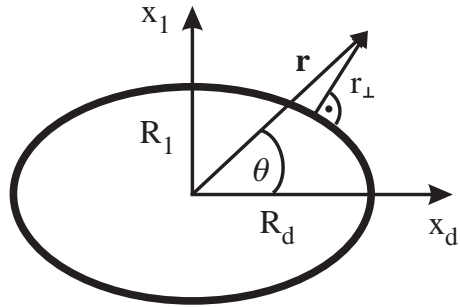}
\caption{Schematic side view of an ellipsoid with semi-axes 
$R\equiv R_1 \le R_2 \le \cdots \le R_d$. Only the projection of the particle onto the 
$x_1-x_d$ plane is shown. Any point ${\bf r}$ outside the particle 
can be reached by a distance $r_\perp$ measured from the point closest to  
${\bf r}$ on the surface of the particle. The angle between the position vector of 
the latter surface point and the $x_d$ axis is denoted as $\theta\equiv\theta_d$.}
\label{fig1}
\end{center}
\end{figure}

In a classical binary liquid mixture near its critical demixing point, the 
order parameter is a suitable concentration difference between the two species 
forming the liquid. The generic preference of confining boundaries for one of the 
two species 
results in the presence of effective surface fields leading to nonvanishing order 
parameter profiles even in the one-phase region of the phase diagram
\cite{fish:78,fish:81}. These critical adsorption profiles become particularly 
long-ranged due to correlation effects induced by the critical fluctuations 
of the order parameter of the solvent. While several theoretical investigations 
have been devoted to the understanding of critical adsorption phenomena on planar 
walls and spherical particles (see Refs.~\cite{law:01,schl:03} and references therein), 
critical adsorption on non-spherical particles has been studied only in the 
limiting cases of infinitely long cylinders \cite{hank:99a} and very small 
dumbbells and ellipsoids within the framework of a small particle operator 
expansion \cite{eise:04}, despite the growing interest in non-spherical 
colloidal particles (see, e.g., Refs.~\cite{harn:07,dogi:06}). Due to the reduced
symmetry of the shapes of non-spherical colloidal particles compared with spherical 
symmetry the interactions between them depend not only on the separation between 
their centers but also on their mutual orientations. As a result various properties 
of fluids consisting of such particles differ from the corresponding ones of fluids 
consisting of spherical particles. As a prerequisite for studies -- motivated by the 
aforementioned aggregation phenomena near criticality -- on critical Casimir 
forces between non-spherical colloidal particles and surfaces we investigate 
here systematically the temperature dependence of critical adsorption on a 
single colloidal particle. In order to treat ellipsoids, spheres, 
and cylinders in a unified way within the appropriate field-theoretical approach 
and for general embedding dimensions $D$ it is helpful to consider the particle 
shape of a hypercylinder defined as
%
%
\begin{multline}
{\cal K}_d(\{R_i\}) = \Bigg\{{\bf r}=(x_1, x_2, \cdots, x_D)\, \\
\left| \hspace*{0.2cm}\sum\limits_{s=1}^d\left(\frac{x_s}{R_s}\right)^2\right.
\le 1\,,\hspace*{0.2cm}d\le D\Bigg\}\,,
\end{multline}
%
%
where $R_1 \le R_2 \le \cdots \le R_d$ are the semi-axes of the particle
(see Fig.~\ref{fig1}). In the case of equal semi-axes $R_1=R_2=\cdots =R_d$ 
the hypercylinder reduces to the so called generalized cylinder with an 
infinitely extended ''axis'' of dimension $D-d$ \cite{eise:96,hank:99a}.
In $D=3$ the ''axis'' can be the axis of an ordinary infinitely elongated 
cylinder ($d=2$), or the midplane of slab ($d=1$), or the center of a sphere 
($d=3$). In the case that not all semi-axes are equal but $d=D$ the hypercylinder
reduces to an ellipsoid which is called a spheroid if the lengths of two 
semi-axes are the same. In $D=3$ one may distinguish prolate spheroids
($R_1=R_2<R_3$) from oblate spheroids ($R_1<R_2=R_3$). The general ellipsoid 
($R_1<R_2<R_3$) is called a triaxial ellipsoid.
The generalization of $D$ to values different from three 
is introduced for technical reasons because $D_{uc}=4$ is the upper critical 
dimension for the relevance of fluctuations of the order parameter leading to 
a behavior different from that obtained from mean field theory valid in $D=4$.
It proves convenient to express the position vector ${\bf r}$ in terms of a 
distance $r_\perp$ perpendicular to the surface and dimensionless angles 
$\{\theta_s\}$, $s=2, \cdots, D$ (see Fig.~\ref{fig1}). Moreover, the shape of a 
hypercylinder can be characterized by the smallest semi-axis $R\equiv R_1$ and the 
dimensionless ratios $\{\delta_s=R_s/R\}$, $s=2, \cdots, d$.

\section{Order parameter profiles}
\subsection{Scaling properties and short distance expansion} \label{shortdist}
Close to $T_c$ the critical adsorption on the surface of the mesoscopic particle
is characterized by an order parameter profile $\langle\phi({\bf r})\rangle_t$ 
which takes the following scaling form
\begin{equation} \label{eq2}
\langle\phi({\bf r})\rangle_{t}=a|t|^\beta 
P_\pm(r_\perp/\xi_\pm,R/\xi_\pm,\{\theta_s\},\{\delta_s\})\,
\end{equation}
for distances $r_\perp$ from the surface sufficiently larger than a typical 
microscopic length. Here $\langle\cdots\rangle_{t}$ denotes the thermal 
average. $\xi_\pm(t\to 0)=\xi_0^\pm |t|^{-\nu}$ is the bulk correlation
length above ($+$) or below ($-$) the critical temperature $T_c$, where 
$t=(T-T_c)/T_c$ is the reduced temperature and $\beta$ and $\nu$ are the 
standard bulk critical exponents. The scaling functions 
$P_\pm(r_\perp/\xi_\pm,R/\xi_\pm,\{\theta_s\},\{\delta_s\})$ are universal once 
the nonuniversal bulk amplitudes $a$ and $\xi_0^\pm$ are fixed by the value
$\langle\phi_b\rangle_{t\to 0^-}=a|t|^\beta$ of the order parameter in the 
unbounded bulk and by the true correlation length defined 
by the exponential decay of the bulk two-point correlation function in real 
space. Therefore on finds 
\mbox{$P_-(r_\perp/\xi_-\to \infty,R/\xi_-,\{\theta_s\},\{\delta_s\})=1$}
and 
\mbox{$P_+(r_\perp/\xi_+\to \infty,R/\xi_+,\{\theta_s\},\{\delta_s\})=0$}.
In the opposite limit $r_\perp/\xi_\pm\to 0$, i.e., $T\to T_c$, the scaling 
functions and the order parameter profile exhibit short-distance singularities 
in the form of power laws which reflect the anomalous scaling dimension of the 
order parameter:
\begin{equation} \label{eq3}
\langle\phi({\bf r})\rangle_{t=0}=a
C_\pm(r_\perp/R,\{\theta_s\},\{\delta_s\})
\left(\frac{r_\perp}{\xi_0^\pm}\right)^{-\beta/\nu}\,.
\end{equation}
The ratio $\beta/\nu$ of the critical exponents has the value 1 in $D=4$,
$\approx 0.517$ in $D=3$ \cite{guid:98}, and $1/8$ in $D=2$.
The amplitude functions $C_\pm(r_\perp/R,\{\theta_s\},\{\delta_s\})$ are 
universal but depend on the definition of the correlation length, as 
the scaling functions 
$P_\pm(r_\perp/\xi_\pm,R/\xi_\pm,\{\theta_s\},\{\delta_s\})$, too. Using the 
operator-product expansion (see, e.g., Refs.~\cite{card:86,card:90}) a short 
distance expansion of the universal amplitude functions can be derived:
%
%
\begin{multline}
\label{eq4}
\frac{C_\pm(r_\perp/R\to 0,\{\theta_s\},\{\delta_s\})}{c_\pm} = 1+\lambda_{H} H\frac{r_\perp}{R} \\
    +\left[\lambda_{K} K + \lambda_{H^2} H^2\right] \left(\frac{r_\perp}{R}\right)^2 \\
    +\left[\lambda_{H^3} H^3+\lambda_{HK} HK+\lambda_{G} G\right] \left(\frac{r_\perp}{R}\right)^3
    + O\left(\frac{r_\perp}{R}\right)^4\,,
\end{multline}
%
%
where $H\equiv H(\{\theta_s\},\{\delta_s\})$, $K\equiv K(\{\theta_s\},\{\delta_s\})$, 
and   $G\equiv G(\{\theta_s\},\{\delta_s\})$ are dimensionless local 
curvatures characterizing the surface and 
$\lambda_I$ with $I=H, K, H^2, H^3, HK, G$ are dimensionless 
coefficients which depend on $D$ but not on the shape of the 
boundary surface, i.e., they are independent of $\{\theta_s\}$ and 
$\{\delta_s\}$. The dimensionless local curvatures are related to the local 
radii of curvature $A_i\equiv A_i(\{\theta_s\},\{\delta_s\})$ according to 
(see, e.g., Ref.~\cite{davi:89})
%
%
\begin{subequations}
\begin{align}
\label{eq5}
H &= R\sum\limits_{i=1}^{D-1}A_i^{-1}, \quad
K = R^2\sum\limits_{\stackrel{i,j=1}{i<j}}^{D-1}A_i^{-1}A_j^{-1}, \\
G &= R^3\sum\limits_{\stackrel{i,j,k=1}{i<j<k}}^{D-1}
A_i^{-1}A_j^{-1}A_k^{-1}\,. 
\end{align}
\end{subequations}
%
%
In the limit $r_\perp/R\to 0$ the universal amplitude functions 
$C_\pm(r_\perp/R=0,\{\theta_s\},\{\delta_s\})$ reduce to the universal 
amplitudes $c_\pm$ for the critical adsorption profiles near planar walls
confining semi-infinite systems \cite{floe:95}.
According to Eq.~(\ref{eq3}) $C_-(r_\perp/R,\{\theta_s\},\{\delta_s\})=
(\xi_0^+/\xi_0^-)^{\beta/\nu}C_+(r_\perp/R,\{\theta_s\},\{\delta_s\})$, 
where $(\xi_0^+/\xi_0^-)^{\beta/\nu}=c_-/c_+\approx 1.4$ (for $D=3$)
is a universal amplitude ratio which depends on the embedding dimension $D$ 
(see Table I in Ref.~\cite{hank:99a}). 
Therefore it is sufficient to consider only one of these functions.

In the case of a sphere, i.e., $H^{(sph)}=D-1$, $K^{(sph)}=(D-1)(D-2)/2$,
and $G^{(sph)}=(D-1)(D-2)(D-3)/6$, the universal amplitude functions are 
known exactly for any spatial dimension $D$ by means of a finite conformal 
mapping from the half-space \cite{burk:85}:
\begin{eqnarray} \label{eq6}
C^{(sph)}_+(r_\perp/R)=c_+\left(1+\frac{1}{2}\frac{r_\perp}{R}\right)^{-\beta/\nu}\,.
\end{eqnarray}
A comparison of the short distance expansion of $C^{(sph)}_+(r_\perp/R)$
with Eq.~(\ref{eq4}) yields
\begin{eqnarray} \label{eq7}
\lambda_{H}(D-1)=-\frac{1}{2}\frac{\beta}{\nu}
\end{eqnarray}
and \cite{hank:99a}
%
%
\begin{align}
\label{eq8}
(D-1)\left(\frac{\lambda_{K}(D-2)}{2}+\lambda_{H^2}(D-1)\right)=
\frac{1}{8}\frac{\beta}{\nu}\left(\frac{\beta}{\nu}+1\right),
\end{align}
%
%
as well as
%
%
\begin{multline}
\label{eq9}
(D-1)\left(\lambda_{H^3}(D-1)^2 + \frac{\lambda_{HK}(D-1)(D-2)}{2} \right. \\
\left. +\frac{\lambda_{G}(D-2)(D-3)}{6}\right)
=-\frac{1}{48}\frac{\beta}{\nu} \left(2+3\frac{\beta}{\nu}+\frac{\beta^2}{\nu^2}\right)\,.
\end{multline}
%
%
In $D=2$ all coefficients $\lambda_{I}$ with $I=H, H^2, H^3$ can be deduced 
from Eqs.~(\ref{eq7}) - (\ref{eq9}). For $D>2$ a determination of the coefficients 
appearing in  Eqs.~(\ref{eq8}) and (\ref{eq9}) would require the additional knowledge
of universal amplitude functions around differently shaped walls. However, these 
functions are not available beyond the mean field approach discussed in 
Sec.~\ref{meanfield} below.

\subsection{Curvature expansion} \label{curvatureexp}
Recently a curvature expansion of the density profile of a hard sphere fluid 
close to a big convex particle has been proposed and successfully applied to the 
density profile around a hard ellipsoid \cite{koen:05}. The curvature expansion 
separates the properties of the fluid from the geometry of the big particle and 
allows one to determine density profiles in complex geometries based on those 
obtained in much simpler geometric configurations. Assuming that an analogous 
curvature expansion holds for the order parameter profile 
$\langle\phi({\bf r})\rangle_t$ leads to
\begin{multline}
\label{eq10}
\langle\phi({\bf r})\rangle_t = \phi_t^{(p)}(r_\perp)+ \phi_t^{(H)}(r_\perp,R) H \\
+ \phi_t^{(K)}(r_\perp,R) K \phi_t^{(H^2)}(r_\perp,R) H^2 
+\phi_t^{(H^3)}(r_\perp,R) H^3 \\ +\phi_t^{(HK)}(r_\perp,R) HK + \phi_t^{(G)}(r_\perp,R) G +O(R^{-4}),
\end{multline}
%
%
where $\phi_t^{(p)}(r_\perp)$ is the order parameter profile for the half-space 
bounded by a planar wall and $\phi_t^{(I)}(r_\perp,R)$ with $I=H, K, H^2, HK, H^3, G$ 
are expansion coefficient functions. A comparison of this equation with 
Eqs.~(\ref{eq3}) and (\ref{eq4}) leads to the following short distance 
behavior of the curvature expansion coefficient functions:
%
%
\begin{align}
\label{eq11}
\phi_{t}^{(p)}(r_\perp) =& a c_\pm \left(\frac{r_\perp}{\xi^\pm_0}\right)^{-\beta/\nu} \nonumber
    \psi^{(p)}_\pm(r_\perp/\xi_\pm), \\ 
&\psi^{(p)}_\pm(0) = 1, \\
\phi_{t}^{(I)}(r_\perp,R) = & a c_\pm \left(\frac{r_\perp}{\xi^\pm_0}\right)^{-\beta/\nu}
    \left(\frac{r_\perp}{R}\right)^n \lambda_I \psi^{(I)}_\pm(r_\perp/\xi_\pm), \nonumber \\
&\psi^{(I)}_\pm(0) = 1, 
\label{eq12}
\end{align}
%
%
where $n=1$ for $I=H$, $n=2$ for $I=K$ and $I=H^2$, $n=3$ for $I=HK$ and $I=H^3$ 
as well as $I=G$. These equations demonstrate the link between the short 
distance expansion of the universal amplitude function [Eq.~(\ref{eq4})]
and the curvature expansion of the order parameter profile [Eq.~(\ref{eq10})]. 
The temperature dependence of the short distance 
expansion coefficient functions $\psi^{(I)}_\pm(r_\perp/\xi_\pm)$ follows 
from simple scaling considerations of the order parameter 
\cite{bray:77,dieh:93} applied to the system under consideration:
%
%
\begin{multline}
\psi^{(I)}_\pm(r_\perp/\xi_\pm\to 0) = 1 + \psi^{(I)}_{1,\pm}\left(\frac{r_\perp}{\xi_\pm}\right)^{1/\nu} \\
    +\psi^{(I)}_{2,\pm}\left(\frac{r_\perp}{\xi_\pm}\right)^{2/\nu}
    + \psi^{(I)}_{D,\pm}\left(\frac{r_\perp}{\xi_\pm}\right)^{D}
    + \cdots \\
= 1 + \psi^{(I)}_{1,\pm}\left(\frac{r_\perp}{\xi^\pm_0}\right)^{1/\nu}|t|
+\psi^{(I)}_{2,\pm}\left(\frac{r_\perp}{\xi^\pm_0}\right)^{2/\nu}t^2 \\
+\psi^{(I)}_{D,\pm}\left(\frac{r_\perp}{\xi^\pm_0}\right)^{D}|t|^{D\nu}+\cdots 
\label{eq13}
\end{multline}
where $I=p, H, K, H^2, HK, H^3, G$. Since the last term in Eq.~(\ref{eq13}) 
represents the leading non-analytic contribution to the order parameter profile 
for $t\to 0$, one has 
$\psi^{(I)}_{1,-} (\xi_0^+)^{1/\nu}=-\psi^{(I)}_{1,+} (\xi_0^-)^{1/\nu}$ and
$\psi^{(I)}_{2,-} (\xi_0^+)^{2/\nu}=\psi^{(I)}_{2,+} (\xi_0^-)^{2/\nu}$.

\section{Mean field theory} \label{meanfield}
Within a field theoretical renormalization group approach the scaling functions 
introduced above can be determined in lowest order perturbation theory within the 
framework of mean field theory corresponding to $\epsilon=4-D=0$. A renormalization 
group enhanced mean field theory is obtained by using for the scaling 
variables entering into scaling functions their full scaling form 
for $\epsilon=1$. The standard Ginzburg-Landau fixed-point Hamiltonian for describing 
critical phenomena is given by \cite{bind:83,dieh:86}
\begin{equation} \label{eq14}
{\cal H}[\phi]=\int_V dV\,\left(\frac{1}{2}(\nabla \phi)^2+\frac{\tau}{2}\phi^2+
\frac{u}{24}\phi^4\right)\,.
\end{equation}
This Hamiltonian has to be supplemented by the boundary condition $\phi=+\infty$ 
at the surface of the colloidal particle corresponding to the critical adsorption 
fixed point \cite{burk:94}. The integration runs over the volume accessible 
to the fluid and the parameter $\tau$ is proportional to the reduced temperature
$t$. Within mean field theory $\tau=t/(\xi_0^+)^2$ for $T>T_c$ and 
$\tau=-|t|/(\xi_0^-)^2/2$ for $T<T_c$ with $\xi_0^+/\xi_0^-=\sqrt{2}$. The 
coupling constant $u>0$ stabilizes the Hamiltonian ${\cal H}[\phi]$ for temperatures 
below the critical point ($T<T_c$) and $(\nabla \phi)^2$ penalizes spatial 
variations of the order parameter. Within mean field theory fluctuations 
of the order parameter are neglected and only the configuration of the order 
parameter with the largest statistical weight 
$m=\sqrt{u/6}\langle \phi \rangle$ with $\sqrt{u/6}=\sqrt{|\tau|/|t|}/a$ 
is taken into account. After functional minimization one obtains the 
Euler-Lagrange equation
\begin{equation} \label{eq15}
\Delta m=\tau m +m^3\,.
\end{equation}
Equation (\ref{eq15}) can be solved numerically for arbitrary temperatures. 
For computational purposes it is convenient to choose a spheroidal coordinate 
system in which the surface of the ellipsoid corresponds to a constant value of 
one spheroidal coordinate (see, e.g., Ref.~\cite{abra:72}). In the case of a 
generalized cylinder the Euler-Lagrange equation reduces to \cite{hank:99a}
\begin{multline}
\label{eq16}
\frac{\partial^2}{\partial r_\perp^2}m(r_\perp,R,\tau)
+\frac{d-1}{r_\perp+R}
\frac{\partial}{\partial r_\perp}m(r_\perp,R,\tau) \\
=\tau m(r_\perp,R,\tau) +m^3(r_\perp,R,\tau)\,,
\end{multline}
%
%
where $d=1, 2, 3,  4$ denotes different types of generalized cylinders. Using the 
short distance expansion [Eq.~(\ref{eq4})] for the universal amplitude function 
$C_+(r_\perp/R)=r_\perp m(r_\perp,R,\tau=0)$ [Eq.~(\ref{eq3})] as input into 
Eq.~(\ref{eq16}) and equating terms of the same power in $r_\perp/R$ leads to 
%
%
\begin{align}
\label{eq17}
\lambda_H=-\frac{1}{6}, \quad \lambda_K=-\frac{1}{3}, \quad \lambda_{H^2}=\frac{5}{36}, \nonumber \\
\lambda_{H^3}=-\frac{31}{216}, \quad \lambda_{HK}=\frac{1}{2}, \quad \lambda_{G}=-\frac{3}{4},
\end{align}
%
%
where $H=d-1$, $K=(d-1)(d-2)/2$, and $G=(d-1)(d-2)(d-3)/6$.

\subsection{Curvature expansion coefficient functions}

The curvature expansion coefficient functions in Eq.~(\ref{eq10}) are determined by 
solving Eq.~(\ref{eq16}) for the four generalized cylinders with high symmetries
and for arbitrary values of $\tau$. In the following we restrict our presentation 
to the case $T>T_c$. The first curvature expansion coefficient function is
known analytically \cite{hank:99a} (see also the Appendix):
\begin{eqnarray} \label{eq18}
\psi^{(p)}_+(\zeta_+=r_\perp/\xi_+)=\frac{\zeta_+}{\sinh(\zeta_+)}\,.
\end{eqnarray}

\begin{figure}
\begin{center}
\includegraphics[ width=6.5cm]{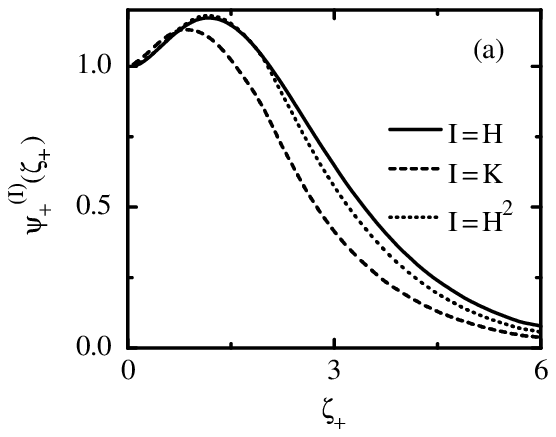}\\[35pt]
\includegraphics[ width=6.5cm]{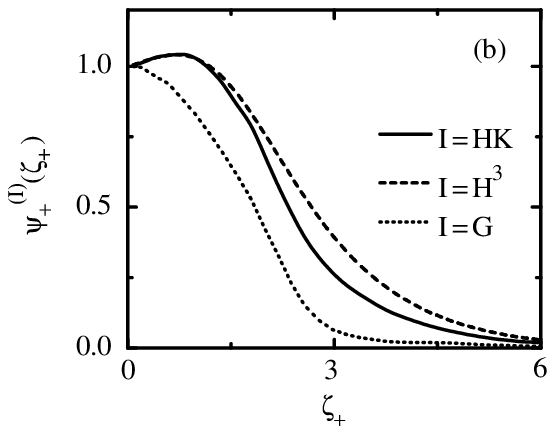}
\caption{(a) [(b)] The curvature expansion coefficient functions 
$\psi^{(I)}_+(\zeta_+=r_\perp/\xi_+)$ for $I=H, K, H^2$ 
[$HK, H^3, G$] of the order parameter profile for critical adsorption on curved 
surfaces according to Eqs.~(\ref{eq12}) and (\ref{eq13}).
The coefficient functions
have been determined from the numerical solution of Eq.~(\ref{eq16}) for 
$d=2, 3, 4$ corresponding to three different types of generalized cylinders 
with high symmetries in spatial dimension $D=4$. All coefficient functions 
attain the finite value 1  as  $\zeta_+\to 0$ (see Eq.~(\ref{eq13})).} \label{fig2}
\end{center}
\end{figure}

Our numerical data obtained from Eq.~(\ref{eq16}) for $d=2, 3, 4$ can be used 
to determine individually
all coefficient functions appearing in Eq.~(\ref{eq10}). Moreover, 
the internal consistency of Eq.~(\ref{eq10}) can be checked in the case of the 
functions $\phi_t^{(H)}(r_\perp,R)$, $\phi_t^{(K)}(r_\perp,R)$, and 
$\phi_t^{(H^2)}(r_\perp,R)$ in addition to the function $\phi_t^{(p)}(r_\perp)$ 
for the half-space. This means that first each of these functions is determined 
separately for each value of $d$, e.g., $\phi_{t,d=2}^{(H)}(r_\perp,R)$,
$\phi_{t,d=3}^{(H)}(r_\perp,R)$, and $\phi_{t,d=4}^{(H)}(r_\perp,R)$. 
Thereafter these three functions as obtained this way are compared with each 
other. Internal consistency means that these three functions
are identical within the numerical accuracy.
For higher orders $n\ge 3$ in Eq.~(\ref{eq12}), there 
are contributions of more than two coefficient functions to the curvature 
expansion. A consistency check for these functions would require 
the additional knowledge of the order parameter profiles around differently 
shaped surfaces. For comparison we note that in the case of the aforementioned 
curvature expansion of density profiles of a hard sphere fluid around an ellipsoid 
in $D=3$ it was possible to check the internal consistency of the curvature
expansion only in the case of the density profiles corresponding to 
$\phi_t^{(H)}(r_\perp,R)$ and $\phi_t^{(p)}(r_\perp)$ \cite{koen:05}.
Hence the present calculation in $D=4$ provides a more stringent consistency test. 
Figures \ref{fig2} (a) and (b) display the curvature expansion coefficient functions
except for the well-known function $\psi^{(p)}_+(\zeta_+=r_\perp/\xi_+)$ for the 
half-space [Eq.~(\ref{eq18})]. One observes that all coefficient functions attain the 
contact value 1 at $\zeta_+=0$ due to 
$\psi^{(I)}_+(\zeta_+\to 0)-1=a_+^{(I)} (\zeta_+)^{1/\nu}$ 
in agreement with Eq.~(\ref{eq13}), where $\nu=1/2$ within mean field theory
and $a_+^{(I)}$ is a dimensionless coefficient (see the Appendix).
Moreover, all coefficient functions exhibit an exponential decay for large values 
of the scaling variable $\zeta_+=r_\perp/\xi_+$. We have verified numerically the 
internal consistency of the first four terms of the 
curvature expansion [Eq.~(\ref{eq10})] as discussed above.

\subsection{Scaling functions}

\begin{figure}
\begin{center}
\includegraphics[ width=6.5cm]{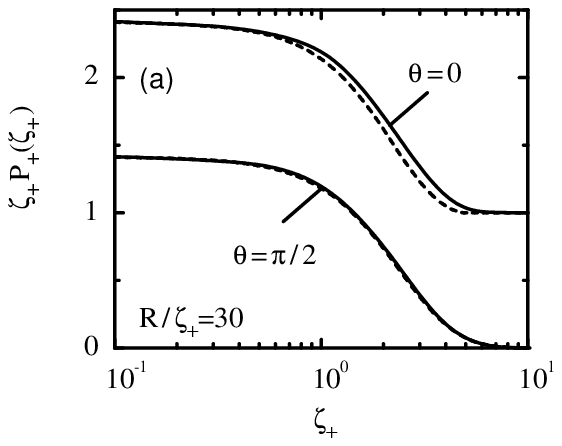}\\[35pt]
\includegraphics[ width=6.5cm]{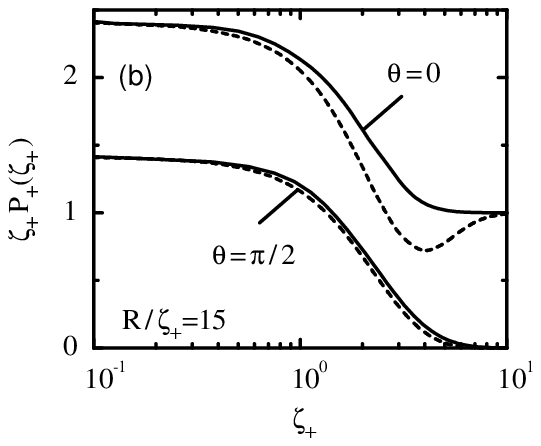}
\caption{(a) [(b)] The scaling function
$P_+(\zeta_+=r_\perp/\xi_+,R/\xi_+,\{\theta_s\},\{\delta_s=R_s/R\})$ [Eq.~(\ref{eq2})]
of the order parameter profile for critical adsorption on an ellipsoid 
(see Fig.~\ref{fig1}) with $R/\xi_+=30$ [$R/\xi_+=15$] and $\delta_2=\delta_3=\delta_4/2=1$ 
in dimension $D=4$ for the angles $\theta=\theta_4=\pi/2$ (lower curves) and 
$\theta=\theta_4=0$ (upper curves). For symmetry reasons the order parameter 
profile is independent of the angles $\theta_2$ and $\theta_3$.
Direct numerical calculations (solid lines) according to Eq.~(\ref{eq15})  are 
compared with the curvature expansion up to and including third order (dashed lines) 
as obtained from Eq.~(\ref{eq10}). The upper two curves have been shifted up by 1.} 
\label{fig3}
\end{center}
\end{figure} 

We now turn our attention to the order parameter profiles around the ellipsoidal 
particle shown in Fig.~\ref{fig1}. In Figs.~\ref{fig3} (a) and (b) the scaling function 
$P_+=|\tau|^{-1/2}m(\tau)$ of the order parameter profile is shown for two angles 
$\theta$ as obtained from Eq.~(\ref{eq15}) together with the results of the curvature 
expansion [Eq.~(\ref{eq10})]. For $\theta=\pi/2$ and $R/\xi_+=30$
(lower curves in Fig.~\ref{fig3} (a))
the results obtained from the curvature expansion are in agreement basically 
everywhere with the direct calculations of the scaling function. With increasing 
local curvatures (see Fig.~\ref{fig1}), i.e., for $\theta=0$ and $R/\xi_+=30$
in Fig.~\ref{fig3} (a) (upper curves) as well as for $\theta=0, \pi/2$ and 
$R/\xi_+=15$ in Fig.~\ref{fig3} (b), the third-order curvature expansion 
(i.e., including terms up to $n\le 3$) and the direct 
calculations deviate from each other for scaled distances 
$\zeta_+=r_\perp/\xi_+\gtrsim H, K, G$ because higher order terms in the curvature 
expansion are required. Nevertheless, the calculations support the use of the curvature 
expansion of the order parameter profile as a first approximation for large 
ellipsoids. 

\begin{figure}
\begin{center}
\includegraphics[width=6.5cm]{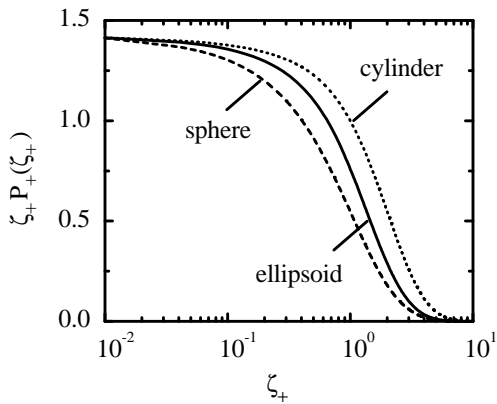}
\caption{The scaling function
$P_+(\zeta_+=r_\perp/\xi_+,R/\xi_+,\{\theta_s\},\{\delta_s=R_s/R\})$ 
(solid line) of the order parameter profile for critical adsorption on an 
ellipsoid with $\delta_2=\delta_3/2=\delta_4/2=1$ and for the angles 
$\theta=\theta_4=\pi/2$ (see Fig.~\ref{fig1}) 
in dimension $D=4$. For symmetry reasons the order parameter 
profile is independent of the angles $\theta_2$ and $\theta_3$.
The dashed and dotted lines represent the results for a 
sphere with equal semi-axes $\delta_2=\delta_3=\delta_4=1$ 
and a generalized cylinder with $\delta_2=1\ll \delta_3=\delta_4$ 
corresponding to $d=4$ and $d=2$ in Eq.~(\ref{eq16}),
respectively. The smallest semi-axis is fixed to $R/\xi_+=0.6$ 
for the ellipsoid, the sphere, and the generalized cylinder.} \label{fig4}
\end{center}
\end{figure} 

\begin{figure}
\begin{center}
\includegraphics[ width=6.5cm]{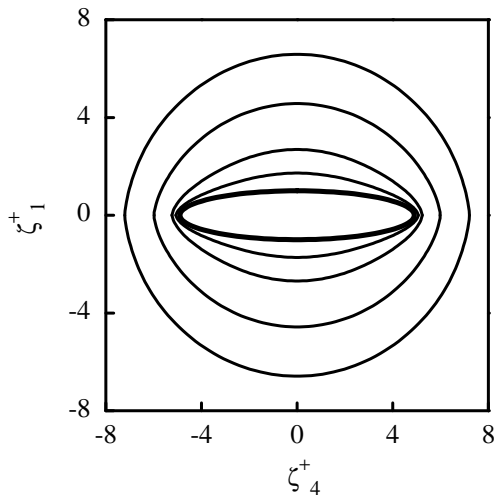}
\caption{The contour lines of the scaling function 
$P_+(\zeta_+=r_\perp/\xi_+,R/\xi_+,\{\theta_s\},\{\delta_s=R_s/R\})$ 
of the order parameter profile for critical adsorption on an ellipsoid (thick line)
in $D=4$ with $R/\xi_+=1$ and $\delta_2=\delta_3=\delta_4=5$. The center of the 
ellipsoid is located at the origin of the coordinate system and only the contour lines in the 
$\zeta_1^+ - \zeta_4^+$ plane are shown, where $\zeta_1^+=x_1/\xi_+$ and 
$\zeta_4^+=x_4/\xi_+$ (see Fig.~\ref{fig1}). The curves correspond to the following 
values of $P_+$: $4, 1, 0.1, 0.01$ (from the middle to the outside).} \label{fig5}
\end{center}
\end{figure} 

The scaling function of the order parameter profile for an ellipsoid with 
two equal small ($R=R_1=R_2$) and two equal large semi-axes ($R_3=R_4$)
is shown in Fig.~\ref{fig4}. In the limit of a large aspect ratio $R_4\gg R$ 
of the ellipsoid the scaling function reduces to the one for 
a generalized cylinder with $d=2$ in Eq.~(\ref{eq16}). In the case that all 
semi-axes are the same the result for a sphere is recovered.
From Fig.~\ref{fig5} one can infer that the contour lines of the scaling 
function of the order parameter profile quickly exhibit spherical symmetry
with increasing distance from the surface of an ellipsoid. Along the 
radial direction $\theta=0$ (i.e., for $\zeta^+_1=x_1/\xi_+=0$) the scaling 
function decays faster away from the surface than along the radial direction 
$\theta=\pi/2$ (i.e., for $\zeta^+_4=x_4/\xi_+=0$) 
because the local curvatures $H$, $K$, and $G$ of the surface at $\theta=0$ 
are larger than at $\theta=\pi/2$ (see Fig.~\ref{fig1}). For comparison 
we note that also in the case of generalized cylinders the decay rate 
of the scaling function increases upon increasing the local curvatures
(see Figs.~5 and 6 in Ref.~\cite{hank:99a}).

\subsection{Universal amplitude functions}

\begin{figure}
\begin{center}
\includegraphics[ width=6.5cm]{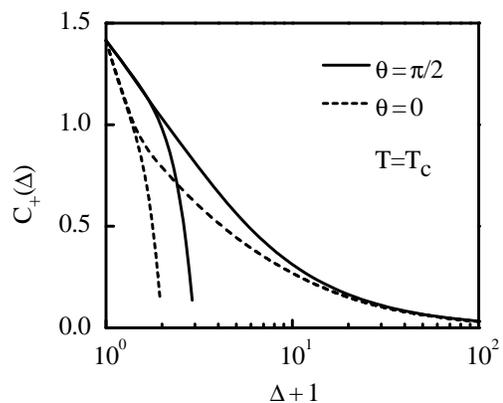} 
\caption{The universal amplitude function 
$C_+(\Delta=r_\perp/R,\{\theta_s\},\{\delta_s=R_s/R\})$ for an
ellipsoid with $\delta_2=\delta_3=\delta_4/2=1$ within mean-field approximation 
(i.e., $D=4$) for the angles $\theta=\theta_4=\pi/2$ and $\theta=0$ 
(see Fig.~\ref{fig1}).
For symmetry reasons the order parameter profile is independent of the angles 
$\theta_2$ and $\theta_3$.
The short lines show the results obtained from the short distance expansion 
(up to and including the order indicated in Eq.~(\ref{eq4})). All curves share the 
common value $C_+(\Delta=0,\{\theta_s\},\{\delta_s=R_s/R\})=c_+=\sqrt{2}$ 
corresponding to the universal amplitude for the half-space.} \label{fig6}
\end{center}
\end{figure} 

\begin{figure}
\begin{center}
\includegraphics[ width=6.5cm]{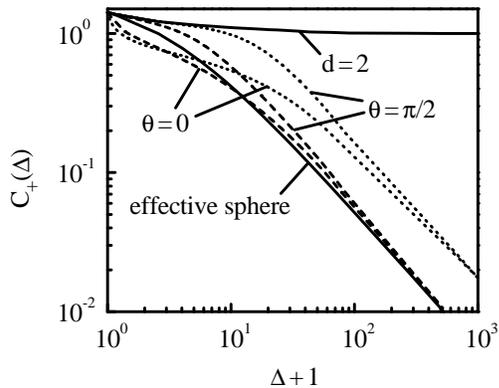} 
\caption{The universal amplitude function 
$C_+(\Delta=r_\perp/R,\{\theta_s\},\{\delta_s=R_s/R\})$ for two
ellipsoids with two equal small and two equal large semi-axes 
$\delta_2=\delta_3/25=\delta_4/25=1$ (dotted lines)
and $\delta_2=\delta_3/4=\delta_4/4=1$ (dashed lines) in mean-field approximation 
(i.e., $D=4$). The upper [lower] dotted and dashed lines represent the results 
for the angle $\theta=\pi/2$ [$\theta=0$] (see Fig.~\ref{fig1}). 
For symmetry reasons the order parameter profile is independent of the angles 
$\theta_2$ and $\theta_3$. The lower 
solid line represents the universal amplitude function for a sphere of radius 
$R_{eff}=R\sqrt{\delta_4^2-1}/\Arcosh(\delta_4)$ with $\delta_4=4$
as obtained from Eq.~(\ref{eq6}). The upper solid 
line follows from Eq.~(\ref{eq16}) with $d=2$ corresponding to a generalized 
cylinder with radius $R$ and two infinitely extended axes, i.e, 
$\delta_2=1$, $\delta_3=\delta_4=\infty$.
All curves share the common value 
$C_+(\Delta=0,\{\theta_s\},\{\delta_s=R_s/R\})=\sqrt{2}$. For large
distances from the surface the universal amplitude function vanishes as 
$\Delta^{-\beta/\nu}$ with $\beta/\nu=1$ in the case of the sphere and the ellipsoids
whereas for the generalized cylinder it tends to the finite \mbox{value $1$.}} 
\label{fig7}
\end{center}
\end{figure} 

Whereas the preceding subsection has been focused on adsorption 
phenomena on ellipsoidal particles for $T\neq T_c$, this subsection 
addresses the adsorption phenomena on ellipsoidal particles in the case 
that the suspending fluid is critical, i.e., at the critical composition and at 
$T=T_c$. The universal amplitude function $C_+(\Delta=r_\perp/R,\{\theta_s\},\{\delta_s\})$ 
(see Eq.~(\ref{eq3})) is shown in Figs.~\ref{fig6} and \ref{fig7} for two different 
ellipsoids. According to Eq.~(\ref{eq4}) the function 
$C_+(\Delta,\{\theta_s\},\{\delta_s\})$ starts 
at $C_+(\Delta=0,\{\theta_s\},\{\delta_s\})=c_+$, where 
$c_+=\sqrt{2}$ within mean field theory. Figure \ref{fig6} shows the 
numerical results as obtained from Eq.~(\ref{eq15}) with 
$C_+=r_\perp m(\tau=0)$ together with the results according to the short distance 
expansion [Eq.~(\ref{eq4})]. The calculations provide the range of 
validity $\Delta=r_\perp/R \lesssim H, K, G$  of the short distance expansion. 
For comparison we note that the curvature 
expansion for $T\neq T_c$ provides a good approximation of the order 
parameter profile for all distances $r_\perp$ from the surface provided 
the local curvatures are smaller than the correlation length 
(see Fig.~\ref{fig3}). However, at the critical point the correlation length
diverges and the curvature expansion, i.e., the short distance expansion, is 
valid only for distances from surface which are small compared with the 
local curvatures of the surface of the colloidal particle.
Figure \ref{fig7} demonstrates that the universal amplitude function decays 
proportional to $r_\perp^{-\beta/\nu}$ with $\beta/\nu=1$ for large distance 
$r_\perp$ from the surface. Far away from the surface of an ellipsoidal 
particle of arbitrary shape the universal amplitude function is angle-independent 
and given by $C^{(sph)}_+(r_\perp/R_{eff})$ [Eq.~(\ref{eq6})] with an effective 
radius $R_{eff}$ that depends on the aspect ratios $\delta_s$ of the ellipsoid
(see the vanishing differences between the dashed lines and the lower solid line 
in Fig.~\ref{fig7} for large $\Delta$). The expression 
$R_{eff}=R\sqrt{\delta_4^2-1}/\Arcosh(\delta_4)$ for the effective radius 
follows from a small particle operator expansion, analogous to the one 
presented in Refs.~\cite{eise:04,eise:03,eise:06}, to the system 
under consideration in this limiting case characterized by semi-axes of the 
ellipsoids which are much smaller than other lengths such as the correlation 
length or the distance between the surface of the particle and the point at which 
the order parameter is monitored.

\section{Summary}

This work has been devoted to the investigation of critical adsorption 
phenomena on ellipsoidal colloidal particles [Fig.~\ref{fig1}] which are 
immersed in a fluid near criticality, i.e., $t=(T-T_c)/T_c \to 0$ in the 
case that the fluid is at the critical composition. The adsorption profiles
are characterized by universal scaling functions $P_\pm$ for $T\neq T_c$
[Eq.~(\ref{eq2})], involving the bulk correlations lengths $\xi_\pm$, 
and for $T=T_c$ by universal amplitude functions $C_\pm$ [Eq.~(\ref{eq3})].

For $T=T_c$ the short distance expansion of the universal amplitude function
of the order parameter profile near a curved surface has been  introduced 
[Eqs.~(\ref{eq3}) and (\ref{eq4})]. This expansion involves local curvatures of 
the surface [Eqs.~(\ref{eq5})]. Also for temperatures $T\neq T_c$ a curvature 
expansion of the order parameter has been considered [Eq.~(\ref{eq10})] which 
involves the local curvatures of the surface, too. The short distance behavior 
of the latter curvature expansion 
coefficient functions [Eqs.~(\ref{eq11}) and (\ref{eq12})] is uniquely determined 
by a comparison of the curvature expansion [Eq.~(\ref{eq10})] with the short 
distance expansion [Eq.~(\ref{eq4})]. A numerical calculation within mean field 
theory [Eq.~(\ref{eq14})] confirms this relation between the curvature expansion
and the short distance expansion [Figs.~\ref{fig2} (a), (b) and \ref{fig6}].

For comparison we note that the contact values of curvature expansion coefficient 
functions of the density profile of a hard sphere fluid close to a hard convex
wall are uniquely determined by exact statistical mechanical sum rules
\cite{hend:86,evan:04} and the morphometric form of the grand potential 
\cite{koen:04,koen:05}. If motion invariance, continuity, and additivity of the  
grand potential are satisfied, only four morphometric measures are needed to 
describe fully the influence of an arbitrarily shaped wall on thermodynamic
properties of fluids consisting of spherical and non-spherical \cite{harn:07} particles.
Hadwiger's theorem \cite{hadw:57,meck:98} states that every motion-invariant,  
continuous, and additive functional in three dimensions can be expressed in terms
of a linear combination of only {\it four} integrated geometric properties:
the volume, the surface area, the mean integrated curvature, and 
the Euler characteristic of the wall. Thermodynamically away from bulk and surface 
phase transitions and for short-ranged interactions and correlations between the 
fluid particles, it is therefore expected that the grand potential fulfills the 
requirements of Hadwiger's theorem. The value of the density profile at a confining 
hard surface can be regarded as a {\it thermodynamic} quantity because of exact 
sum rules implying that the values of the curvature expansion coefficient 
functions of the density profile 
{\it at a confining hard surface}
can be expressed in terms of thermodynamic 
properties such as the pressure of the bulk fluid, the fluid - planar wall
surface tension, and the bending rigidity. These arguments are expected to be not 
applicable to critical phenomena because intrinsic lengths in such systems reach 
macroscopic sizes  so that the assumption of additivity is no longer valid.
This has been confirmed by an explicit calculation of the excess adsorption for 
a sphere and a cylinder immersed in a fluid near criticality 
(see Figs. 8 and 9 in Ref.~\cite{hank:99a}).

However, at $T_c$ and {\it close to a confining surface} 
the curvature expansion coefficient functions 
of the order parameter profile for critical adsorption are determined by the short 
distance expansion of the universal amplitude functions as discussed above;
this expansion involves arbitrarily high orders. 

The curvature expansion for $T\neq T_c$ provides a reliable approximation of 
the order parameter profile for all distances from the surface provided 
the local curvatures are smaller than the inverse correlation length 
[Fig.~\ref{fig3}], similar to the findings for the curvature expansion
of the density profile of a hard sphere fluid close to a hard convex surface.
However, at the critical point the correlation length
diverges and the curvature expansion, which in this case reduces to the 
short distance expansion, is valid only for distances from surface which 
are small compared with the local radii of curvature of the surface of 
the colloidal particle [Fig.~\ref{fig6}]. In this sense at $T_c$
the curvature expansion breaks down as a globally reliable approximation.

The scaling functions for critical adsorption on ellipsoidal particles
of various shapes and sizes have been determined numerically for $T\neq T_c$
within mean field theory [Figs.~\ref{fig3}, \ref{fig4}, and \ref{fig5}]. 
In the case of very small or very large ratios of the semi-axes of the ellipsoid 
the results for a generalized cylinder such as a sphere or a rod are recovered. 
At the surface the explicit form of the universal amplitude function $C_+$ 
for an ellipsoid  [Fig.~\ref{fig7}] attains the value for the half-space 
and exhibits a power law limiting behavior far from the surface in 
accordance with the results for a sphere [Eq.~(\ref{eq6})].

\acknowledgments
The authors thank R. Roth for useful discussions.

\appendix*

\section{}

In this appendix we briefly discuss how the coefficients $\psi^{(I)}_{1,\pm},
\psi^{(I)}_{2,\pm}, \cdots$ for $I=p, H$ introduced in Eq.~(\ref{eq13}) can be 
calculated analytically within mean field theory, i.e., for $D=4$. Since the approaches for 
$T>T_c$ and $T<T_c$ are similar we restrict our presentation to the case $T>T_c$ and 
drop the subscript '+' in order to simplify the notation.

In the case of a generalized cylinder the scaling function 
$P(\zeta=r_\perp/R,\alpha=R/\xi)=|\tau|^{-1/2}m(r_\perp,R,\tau)$ satisfies the 
Euler-Lagrange equation [Eq.~(\ref{eq16})]
\begin{eqnarray} \label{app1}
P''(\zeta,\alpha)+\frac{d-1}{\zeta+\alpha}P'(\zeta,\alpha)
=P(\zeta,\alpha)+P^3(\zeta,\alpha)\,,
\end{eqnarray}
where the derivatives are taken with respect to the variable $\zeta$.
For $\alpha\gg 1$ one may assume that the scaling function is analytic 
in $\alpha^{-1}$ such that it can be expanded into a Taylor series 
around $\alpha^{-1}=0$:
\begin{eqnarray} \label{app2}
P(\zeta,\alpha\to \infty)=P_0(\zeta)+P_1(\zeta)\alpha^{-1}+\cdots\,.
\end{eqnarray}
By using the expansion $(\zeta+\alpha)^{-1}=\alpha^{-1}+\cdots$ in Eq.~(\ref{app1})
in conjunction with Eq.~(\ref{app2}) one obtains the familiar nonlinear 
differential equation for the profile near a planar wall confining a semi-infinite 
system,
\begin{eqnarray} \label{app3}
P_0''(\zeta)=P_0(\zeta)+P_0^3(\zeta)\,,
\end{eqnarray}
with the solution

\begin{eqnarray} \label{app4}
P_0(\zeta)=\frac{\sqrt{2}}{\sinh(\zeta)}=
\frac{P_0^{(1)}}{\zeta}+P_0^{(2)}\zeta+P_0^{(3)}\zeta^3+\cdots
\end{eqnarray}
and the linear differential equation
\begin{eqnarray} \label{app5}
P_1''(\zeta)+(d-1)P_0'(\zeta)=P_1(\zeta)+3P_0^2(\zeta)P_1(\zeta)\,.
\end{eqnarray}
While the profile $P_0(\zeta)$ is well-known we now turn our attention to the calculation 
of $P_1(\zeta)$. To this end we expand $P_1(\zeta)$ as
%
%
\begin{multline}
\label{app6}
P_1(\zeta)=
P_1^{(0)}+P_1^{(1)}\zeta + P_1^{(2)}\zeta^2 + P_1^{(3)}\zeta^3
+ P_1^{(4)}\zeta^4+\cdots\,.
\end{multline}
%
%
By inserting Eqs.~(\ref{app4}) and (\ref{app6}) into Eq.~(\ref{app5}) 
and equating terms with the same power in $\zeta$ one derives the 
coefficients
%
%
\begin{align}
\label{app7}
P_1^{(0)}=-\frac{d-1}{3\sqrt{2}}, \quad P_1^{(1)}=0, \nonumber \\
P_1^{(2)}=-\frac{d-1}{6\sqrt{2}}, \quad P_1^{(4)}=-\frac{d-1}{72\sqrt{2}}.
\end{align}
%
%
However, the value of $P_1^{(3)}$ cannot be determined this way, i.e., 
for {\it any} value of $P_1^{(3)}$ Eq.~(\ref{app5}) is satisfied because of 
the {\it known} values of $P_0^{(1)}, P_0^{(2)}, P_0^{(3)}, \cdots$.
In Ref.~\cite{hank:99a} the following analytic solution of 
Eq.~(\ref{app5}) has been proposed:
%
%
\begin{multline}
\label{app8}
P_1(\zeta) = (d-1) \frac{\cosh(\zeta)-\sinh(\zeta)}{6\sqrt{2}\sinh^2(\zeta)} 
    \Big[ 2\cosh(2\zeta)-2 - 3\zeta \\ - 3\zeta\cosh(2\zeta) +
    3 \sinh(2\zeta)-3\zeta\sinh(2\zeta)\Big].
\end{multline}
%
%
This function can be expanded leading to the coefficients given in 
Eq.~(\ref{app7}) and to
\begin{eqnarray} \label{app9}
P_1^{(3)}=\frac{\sqrt{2}(d-1)}{15}\,.
\end{eqnarray}
By inserting the scaling function $P(\zeta,\alpha)$ from 
Eqs.~(\ref{app2}), (\ref{app6}), and (\ref{app7}) into Eq.~(\ref{eq2})
one obtains the expansion coefficient functions  $\psi^{(I)}_{1},
\psi^{(I)}_{2}, \cdots$ for $I=H$ in Eq.~(\ref{eq13}). We emphasize 
that a non-vanishing value of $P_1^{(3)}$ would lead to an additional 
{\it non-analytic} term $\sim t^{3/2}$ in Eq.~(\ref{eq13}). 
However, according to the scaling considerations in Eq.~(\ref{eq13})
the last term therein $\sim |t|^{D\nu}$ ($=t^2$ within mean field theory)
represents the {\it leading} non-analytic contribution to the 
order parameter profile for $t\to 0$. Therefore one concludes that $P_1^{(3)}=0$
and the analytic expression in Eq.~(\ref{app8}) does not capture fully 
the correct analytic properties of $P_1(\zeta)$. Thus Eq.~(\ref{app5})
has to be solved numerically.

\end{document}